\documentclass[aps,prb,twocolumn]{revtex4}
\usepackage{graphics,amsmath}
\usepackage{tabularx,graphicx}
\usepackage[latin1]{inputenc}
\usepackage{epsfig}

\newcommand{\vf}{v_{\rm F}}

\newcommand{\kf}{k_{\rm F}}

\begin{document}
\title{Electrostatic interactions between graphene layers and their
  environment.}
\author{J. Sabio$^1$}
\author{C. Seo\'anez$^1$}
\author{S. Fratini$^{1,2}$}
\author{F. Guinea$^1$}
\author{A. H. Castro Neto$^3$}
\author{F. Sols$^4$}
\affiliation{$^1$Instituto de Ciencia de Materiales de Madrid
(CSIC), Sor Juana In\'es de la Cruz 3, E-28049 Madrid, Spain \\ $^2$
Institut N\'eel - CNRS \& Universit\'e Joseph Fourier, BP 166,
F-38042 Grenoble Cedex 9, France  \\ $^3$ Department of Physics,
Boston University, 590 Commonwealth Av., Boston, MA 02215, USA
\\ $^4$ Departamento de F{\'\i}sica de Materiales, Universidad
Complutense de Madrid, E-28040 Madrid, Spain.}

\begin{abstract}
We analyze the electrostatic interactions between a single graphene
layer and a SiO$_2$ susbtrate, and other materials which may exist
in its environment. We obtain that the leading effects arise from
the polar modes at the SiO$_2$ surface, and water molecules, which
may form layers between the graphene sheet and the substrate. The
strength of the interactions implies that graphene is pinned to the
substrate at distances greater than a few lattice spacings. The
implications for graphene nanoelectromechanical systems, 
and for the interaction between
graphene and a STM tip are also considered.
\end{abstract}

 \maketitle

\section{Introduction.}

Graphene is a versatile two dimensional material whose singular electronic
and mechanical properties show a great potential for
applications in nanoelectronics.\cite{Netal05b,GN07,NGPNG07} 
Since free floating graphene is subject to crumpling,\cite{Nelson}
the presence of a substrate, and the environment that comes with it, 
is fundamental for its stabilization. 
Hence, this environment will have direct impact
 in the physical properties of graphene. Though the influence of the
substrate and other elements of the 
surroundings has been taken into account in different ways in the
literature, the exact part that these are playing
is not yet fully understood.
  
On the one hand,  the 
differences observed between samples grown on different substrates
constitute an open issue. Most
experiments have been carried out in graphene samples deposited over SiO$_2$, 
or grown over SiC substrates, \cite{Betal04} and a
better understanding of how graphene properties are expected to change
would be worthy. On the other hand, there is the question of
characterizing all the effects that a particular environment has on
graphene electronic and structural properties. 

Concerning electronic
properties, it has been suggested that the low temperature
mobility of the carriers is determined by scattering with charged impurities
in the SiO$_2$ substrate,\cite{NM07,AHGS07} and the effect of these charges
can be significantly modified by the presence of water
molecules.\cite{Setal07b} Actually, the very polar modes of
SiO$_2$ give a good description of the finite temperature 
corrections to the mobility.\cite{PR06,FG07,CJXIF07}
Supporting this idea, recent experiments show that graphene suspended
above the substrate has a higher mobility.\cite{Betal08,XuDu08}

Experiments also seem to reveal a very important role played by the
substrate in the structural properties
of graphene.   STM measurements suggest that single layer graphene follows the
corrugations of the  SiO$_2$ substrate,\cite{ICCFW07,Setal07} and
experiments on graphene nanoelectromechanical systems (NEMS) 
indicate that the  substrate induces
significant stresses in few layer graphene samples.\cite{Betal07}
Moreover, the interaction between graphene and the substrate determines
the frequency of the out of plane (flexural) vibrations, which can
influence the transport properties at finite
temperatures.\cite{KG07,MNKSEJG07}

\begin{figure}
\begin{center}
\includegraphics[width=8.5cm]{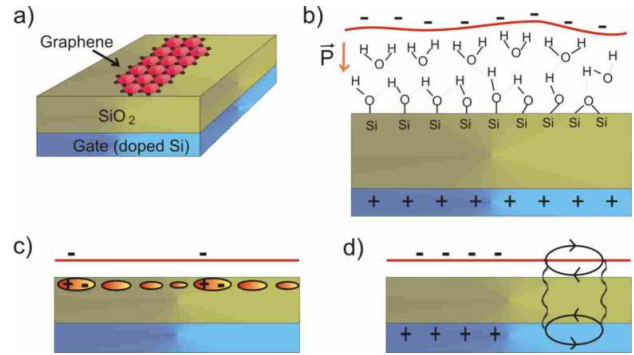}
\end{center}
\caption{a) Sketch of the system studied in the text. Interaction
effects: b) Interaction with water molecules attached to hydroxyl
radicals at the substrate. c) Interaction with polar modes at the
surface of the substrate. d) van der Waals interaction between the
graphene sheet and the metallic gate.} \label{mechanisms}
\end{figure}

In order to shed light on the influence of the environment on the
graphene properties,  we 
analyze the characteristic energies of 
interaction with the substrate and other
materials present in the experimental setup. This allows us to
evaluate the relative importance of the different interactions 
in  the binding and mechanical response of the graphene layer. We also
provide estimates of quantities
such as equilibrium distances, typical lengthscales of corrugations, and
frequencies of vibration,
which can be measured in principle in current experimental setups.

Throughout the paper we
concentrate on SiO$_2$, though results are easily generalized to other
substrates. Particularly,  we consider: i)  the van der
Waals forces between graphene and the metallic gate below the
SiO$_2$ substrate, ii) the electrostatic forces between the graphene
layer and the polar modes of the substrate, iii) the electrostatic
forces between  graphene and charged impurities  which may be
present within the substrate and iv) the electrostatic forces
between  graphene and a water layer which may lay between graphene
and the substrate.\cite{Setal07b}  A sketch of the setup studied,
and the different interaction mechanisms, is shown in
Fig.\ref{mechanisms}. We will also mention the possibility of weak
chemical bonds between the graphene layer and molecules adjacent to
it,\cite{LPP07,Wetal07} although they will not be analyzed in
detail. We do not consider a possible chemical modification of the
graphene layer,\cite{Eetal07,Wetal07b} which would change its transport
properties.

The general features of the electrostatic interactions to be studied
are discussed in the next section. Then, we analyze, case by case,
the different interactions between the graphene layer and the
materials in its environment. Section IV discusses the main
implications for the structure and dynamics of graphene, with
applications to graphene NEMS and the interaction between graphene
and a STM tip. The last section presents the main highlights of our
work.

\section{Electrostatic interactions between a graphene layer and its
environment.} 
The electrons in the $\pi$ and $\pi^*$ bands  of
graphene are polarized by electromagnetic potentials arising from
charges surrounding it.  The van der Waals interactions between
metallic systems, and metals
and graphene can be expressed as integrals over the dynamic
polarizability of both systems. Those, in turn, can be written in
terms of the zero point energy of the plasmons.\cite{TA83,DWR06} The
interaction between the graphene layer and a polarizable dielectric
like SiO$_2$ is also given by an integral of the polarizability of
the graphene layer times the polarizability of the dielectric. The latter
can be approximated by the propagator of the polar modes, which play
a similar role to the plasmons in a metal. The interaction between
the graphene and static charges of electric dipoles depends  only on the
static polarizability.\cite{image}

\begin{figure}
\begin{center}
\includegraphics*[width=6cm]{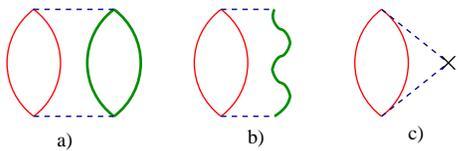}
\end{center}
\caption{Lowest order diagrams which contribute to the interaction
between: a) graphene and a metal, b) graphene and a polar
dielectric, and c) graphene and a static charge distribution. The
thin red bubble stands for the graphene susceptibility. The thick
green bubble represents the metallic susceptibility. The wavy green
line stands for the propagator of a phonon mode in the dielectric.
Crosses stand for static charge distributions, and dashed lines represent the
electrostatic potential.} \label{diagrams}
\end{figure}

We will calculate these interactions using second order perturbation
theory, assuming a perfect graphene sheet so that
the momentum parallel to it is conserved.
The corresponding diagrams are given in
Fig. \ref{diagrams}. All interactions depend, to this order,
linearly on the polarizability of the graphene layer. In ordinary
metallic systems, the Coulomb interaction is changed qualitatively
when screening by the graphene electrons is taken into account through
an RPA summation of diagrams. This is not the
case for undoped graphene. There, 
the Random Phase Approximation leads to a finite
correction  $
\pi e^2 / 8 \hbar \vf \sim 1$ to the dielectric constant, which
 does not change significantly the estimates obtained
using second order perturbation theory.

The response function of a
graphene layer at half filling is:\cite{GGV94}
\begin{equation}
\chi_G (\vec{q}, i\omega) = \frac{N_v N_s}{16 \hbar}
\frac{q^2}{\sqrt{v_F^2 q^2 + \omega^2}}, \label{susc}
\end{equation}
where $N_s = N_v = 2$ are the valley and spin degeneracy. This
expression is obtained assuming a linear dispersion around the $K$
and $K'$ points of the Brillouin Zone. It is valid up to a cutoff in
momentum $\Lambda \sim a^{-1}$ and energy $\omega_c \sim \vf
\Lambda$, where $a$ is  the lattice spacing. Beyond this scale, the
susceptibility has a more complex form, and it is influenced by the
trigonal warping of the bands. The component of the electrostatic
potential induced by a system at distance $z$ from the graphene
layer with momentum $\vec{q}$ is suppressed by a factor $e^{- |
\vec{q} | z}$. Hence, the integrations over $\vec{q}$ can be
restricted to the region $0 \le q = | \vec{q} | \lesssim q_{max}
\sim z^{-1}$. The combination of a term proportional to $e^{- |
\vec{q} | z}$ and scale invariant quantities such as the
susceptibility in Eq. (\ref{susc}) leads to interaction energies
which depend as a power law on $z$. In general, we will consider
only the leading term, neglecting higher order
corrections.\cite{polar}

The calculation described above, which is valid for a single
graphene layer at half filling, can be extended to other fillings
and to systems with more than one layer. In all cases, the
calculations are formally the same, and the interaction energies can
be written as integrals over energies and momenta of the
susceptibility of the system being considered, which replaces the
susceptibility of a single layer, Eq. (\ref{susc}). The
susceptibility of a doped single layer is well approximated by that
of an undoped system, Eq. (\ref{susc}), for momenta such that $q
\gtrsim \kf$.\cite{WSSG06} Analogously, the susceptibilities of a
stack of decoupled layers of graphene and multilayered graphene
become similar for $q \gtrsim t_\perp / \hbar \vf$\cite{G07}, where
$t_\perp$ is the hopping in the perpendicular direction. The
susceptibility of a single undoped plane of graphene,
Eq.(\ref{susc}) is an increasing function of $q$, so that the
integrals are dominated by the region $q \sim q_{max} \sim z^{-1}$.
Hence, if $q_{max} \gg \kf$ or $q_{max} \gg t_\perp / \hbar \vf$, the
interaction energies do not change appreciably from the estimates
obtained for a single layer. The corrections can be obtained as an
expansion in powers of either $\kf z$, or $( t_\perp z ) / \hbar \vf$.
The expression  Eq. (\ref{susc}) can  therefore be
considered as the lowest order expansion in these parameters. 
For $z \sim 1$nm, $t_\perp\sim 0.35 eV$  
and carrier densities such that $n
\sim 10^{10} - 10^{12}$ cm$^{-2}$, we obtain $\kf z \sim 10^{-2} -
10^{-1}$ and $ t_\perp / \hbar \vf\sim  10^{-2} -
10^{-1}$. In the following, we will analyze mostly the interaction
energies using the expression in Eq. (\ref{susc}) for the graphene
polarizability.

\section{Interactions with specific environments.}

\subsection{Metallic gate.}
We describe the metallic gate as doped Si, separated from the
graphene layer by a 300nm thick slab of SiO$_2$ dielectric.
For the Si doping and voltages applied, most of the charge in the Si gate
 is concentrated on a layer of about $10$ nm thickness,\cite{S81} much narrower
than the distance to the graphene sheet, so that the gate is
effectively two dimensional. We describe the susceptibility of the
gate as that of a dirty two dimensional electron gas:
\begin{equation}
\chi_{gate}(\vec{q},i\omega) = -\frac{dn}{d\mu} \frac{Dq^2}{Dq^2 +
|\omega|},
\end{equation}where $D = {v_{\rm F}}_{gate} l_{gate}$ is
the diffusion coefficient of the electrons in  the gate,
${v_{\rm F}}_{gate}$ is the
Fermi velocity, $l_{gate}$ is the mean free path, and
$dn / d \mu$ is the bare compressibility, given by the density of
states at the Fermi level (see, for instance, Ref. \onlinecite{GJS04}).

The interaction between the graphene layer and the gate is given by:
\begin{equation}
v_q(z) = \frac{2 \pi e^2}{ \epsilon} \frac{e^{-qz}}{q},
\end{equation}
being $\epsilon$ the static dielectric constant of the SiO$_2$
substrate.

The lowest order contribution to the energy in perturbation theory
has the following form:
\begin{equation}
E_{gate}^{(2)} = - \hbar \sum_q \int_0^{\infty} \frac{d\omega}{2\pi}
v_q^2(z) \chi_G (\vec{q},i\omega) \chi_{gate} (\vec{q},i\omega).
\end{equation}
For future reference, note that we use the symbol $E$ for energies per
unit area, and $\mathcal{E}$ for total (integrated) energies.
The resulting integrals can be calculated analytically in the limit
$z_s\equiv D/4v_F\ll z$:
\begin{equation}
E_{gate}^{(2)} = - \frac{1}{12}\frac{dn}{d\mu} \frac{D}{v_F}
\frac{e^4}{\epsilon^2} \frac{1}{(2z)^3} \left[ \log
\left( \frac{z}{z_s} \right) + \frac{1}{3} \right].
\end{equation}
The dependence on $z^{-3} \log (z/z_s)$ was obtained in Ref.
\onlinecite{DWR06}.

We take, as representative parameters for the gate and the graphene
layer, $D \approx  10^{-3}$ m$^2$/s, $\vf = 10^6$ m/s, $z = 300$
nm, $dn/d\mu \simeq g(E_F) = 0.04$ eV$^{-1}$ \AA$^{-2}$
and $\epsilon = 4$ for the SiO$_2$ substrate.
These parameters lead to interaction energies of order $\sim
10^{-8}$ meV \AA$^{-2}$.

\subsection{Polar dielectric.}

The interaction between the graphene layer and the SiO$_2$ substrate
can be expressed in terms of the electric fields induced by the
surface polar modes of SiO$_2$.\cite{S72,WM72,MA89,Hetal06} The
coupling can be written as:
\begin{equation}
H_I = \sum_q M_q \rho_q \left(b_q + b_{-q}^{\dagger} \right),
\end{equation}
where $\rho_q$ is the electron density operator and
$b_q^{\dagger}$,$b_q$ the creation/destruction operators for
phonons,
and $M_q^2 = (\hbar^2 v_F^2) g e^{-2q z}/(qa)$ is the interaction matrix
element, with $g$ a dimensionless coupling constant.
In SiO$_2$ we have two dominant phonon modes at $\hbar \Omega_{1}=59$ meV and
$\hbar \Omega_{2}=155$ meV, with  $g_1 =
5.4\cdot 10^{-3}$ and $g_2 = 3.5 \cdot 10^{-2}$ respectively.\cite{FG07}

The lowest order contribution to the energy is given by:
\begin{equation}
E^{(2)}_{subs} = \sum_i \sum_{\vec{q}} \int
\frac{d\omega}{2\pi} \chi_G (\vec{q}, i\omega) |M_q(z)|^2
D^{(0)}_i(\vec{q},i\omega),
\end{equation}
where we have introduced the free phonon propagators:
\begin{equation}
D_i^{(0)}(\vec{q},i\omega) = - \frac{2 \Omega_i}{\omega^2 +
\Omega_i^2}.
\end{equation}

The calculation can be again carried out analytically. In the limit $z \ll l_i
\equiv \vf / \Omega_i$ we obtain:
\begin{equation}
E_{subs}^{(2)} = - \sum_i \frac{\hbar v_F}{a}
\frac{g_i}{(2z)^2},
\end{equation}
which gives a $z^{-2}$ dependence on the distance. In the opposite limit $z
\gg l_i$, which can be of interest in suspended graphene experiments, one
obtains:
\begin{equation}
E_{subs}^{(2)} = - \frac{\hbar v_F}{6 a} \frac{1}{(2z)^3}\sum_i l_i g_i 
\left[\log{\frac{l_i}{4 z}} + \frac{1}{3}\right]
\end{equation}
Let us give some numerical estimates for both expressions. In the case of graphene deposited over the substrate, we have $z \sim 1$ nm  (see Ref. \onlinecite{GN07}) 
and  $l_i \ll z$, having interaction
energies of order $E^{(2)}_{subs} \sim - 4\times 10^{-1}$meV
\AA$^{-2}$. For suspended graphene, $z\sim 300$ nm, and the energies are of order $E^{(2)}_{subs} \sim - 10^{-8}$meV
\AA$^{-2}$, i.e., of the same order than the contribution from the gate. 

\subsection{Charges within the substrate.}

In this case the calculations are done considering that effectively
all the charge is concentrated close to the surface of the SiO$_2$
dielectric. The second
order correction to the energy, averaged over the charge distribution, is:
\begin{equation}
E^{(2)}_{ch} = - \sum_{\vec{q}} \chi_G(\vec{q},0) v_q^2(z) n_{imp},
\end{equation}
where we consider a Coulomb interaction $v_q$ between graphene
electrons and charges that is statically screened by the effective
dielectric constant at the interface, $(\epsilon+1)/2$.
Again, this contribution can be carried out
analytically:
\begin{equation}
E^{(2)}_{ch} = - \left(\frac{2e^2}{\epsilon+1}\right)^2
\frac{\pi n_{imp}}{ 2\hbar \vf} \frac{1}{2z}.
\label{En_charges}
\end{equation}
This interaction has a $z^{-1}$ dependence, as the image potential in
ordinary metals. In this case, however, this behavior
arises from the combination of a
vanishing density of states and lack of screening in graphene.

Reasonable values for the impurity concentration in graphene are in
the range $n_{imp} \sim 10^{10} - 10^{12}$
cm$^{-2}$. \cite{NM07,AHGS07} Setting $z\sim 1$nm, typical
interaction energies are of the order $E_{ch}
 \sim - 10^{-4} $-- $10^{-2}$ meV \AA$^{-2}$.

The present result is only valid for graphene samples close to the substrate. 
If the distance to the latter is larger than the typical distance between 
charges, $d_{imp} \sim \sqrt{n_{imp}} \sim 1 - 10 nm$, the electrons 
feel the net effect of the effective charge in the substrate. This is zero in 
average, as there should be a compensated number of positive 
and negative charges. However, if we consider
a finite region of the substrate,  
fluctuations can locally give rise to a net effective charge. This 
can be  estimated by replacing  
$N_{imp} = n_{imp} l^2 \rightarrow \sqrt{n_{imp} l^2}$ 
in the total energy $\mathcal{E}_{ch} = E_{ch}^{(2)} l^2$, where $l$
is the lateral sample size.    

\subsection{Layer of water molecules.}

The properties of the SiO$_2$ surface are dominated, for
thermally grown SiO$_2$ layers, by the presence of abundant silanol
(SiOH) groups, \cite{MM90} whose surface density is about $5 \times
10^{14}$ cm$^{-2}$, unless extra steps like thermal annealing in
high vacuum are taken during the fabrication
process.\cite{SG95,DPX98,N97}
Silanol sites are active centers for
water absorption, so that the SiO$_2$ surface becomes hydrated under
normal conditions,\cite{BV01} which is probably the case of most of
the graphene samples produced by mechanical cleavage.\cite{Setal07b}
Moreover, several layers of water may cover the SiO$_2$ surface, lying
between the oxide surface and the graphene samples after the graphene
deposition. An analogous situation has been shown to happen in experiments
with carbon nanotubes deposited on SiO$_2$. \cite{Ketal03}

The water molecule has an electric dipole, $p_w = 6.2 \times
10^{-30}$ C m $\approx 0.04$e nm. Typical fields applied in present
experimental setups are ${\cal E} \sim 0.1$ V nm$^{-1}$. The energy
of a water dipole when it is aligned with this field is 4 meV
$\sim 50$K, so that, at low temperatures, it will be oriented along the
field direction, perpendicular to the substrate and the graphene
layer.  For this reason, in the following we assume that the water
molecules are not charged, and their dipoles are aligned
perpendicular to the substrate and the graphene layer. This arrangement can
be considered an upper bound to the interaction energy with a neutral water layer, as
inhomogeneities and thermal fluctuations will induce deviations in the
orientation of the dipoles, and will lower the interaction energy. Note,
however, that for high applied electric fields, a charging of water
molecules of the order $Q_{{\rm H_2 O}} \sim 0.1 | e |$ has been
reported.\cite{Ketal03} The presence of these extra charges would
considerably enhance the interaction between the graphene layer and
the water molecules.

A water molecule which is located at a distance $z$ from the
graphene layer induces an electrostatic potential:
\begin{equation}
\Phi ( \vec{q} , z ) = 2 \pi p_w e^{- | \vec{q} | z}.
\end{equation}
This potential polarizes the graphene layer and gives rise to an
interaction energy in a similar way to the static charges discussed
in the preceding section. The lowest order contribution to the
energy is:
\begin{equation}
E^{(2)}_{water} = - \left(e p_w \right)^2
\frac{\pi}{6} \frac{n_w}{\hbar \vf} \frac{1}{(2z)^3},
\label{int_water}
\end{equation}
where $n_w$ is the concentration of water molecules and the $z^{-3}$
behavior arises from the dipolar nature of the interactions.
For $z = 0.3$nm, which is the approximate thickness of a water
monolayer,\cite{AHM01,OSAA07}   the interaction energy is
$E^{(2)}_{water} \sim- 12 n_w$ meV which, for a typical water
concentration $n_w=10^{15}$cm$^{-2}$, yields $E^{(2)}_{water}\sim 1$
meV/\AA$^2$.

The expression in Eq. (\ref{int_water}) can be extended to a
semi-infinite stack of water layers. For simplicity we take
a distance $z$ between graphene
and the uppermost layer of water molecules equal to the interlayer distance.
In this case we obtain:
\begin{equation}
E^{(2)}_{water} = - \left(e p_w \right)^2
\frac{\pi}{6} \frac{n_w}{\hbar \vf} \frac{\zeta(3)}{(2z)^3},
\end{equation}
where $\zeta(3) \approx 1.202$ is Riemann's zeta function. The
present result indicates that the first water layer is the one that
mostly contributes to the binding.

\subsection{Van der Waals interaction between graphene layers.}

For comparison, in this Section we evaluate the van der Waals
interaction between two graphene layers at the equilibrium distance.
Using the same approximations as for the other contributions, we
recover the result of Ref. \onlinecite{DWR06},
\begin{equation}
E_{G-G}^{(2)} = -\frac{\pi e^4}{16 \hbar v_F}  \frac{1}{(2z)^3}.
\end{equation}
For $z = 0.3$nm, this expression gives an interaction energy
of 30 meV \AA$^{-2}$. This estimate is similar to other experimental
and theoretical values of the graphene-graphene
interaction,\cite{Betal98,HNI07} and is at least one
order of magnitude greater than the other contributions analyzed
earlier.

\section{Analysis of the results.}
\subsection{Comparison of the different interactions.}
Numerical estimates for the different interaction energies obtained
for reasonable values of the parameters are listed in
Table \ref{table}.
The present results show that the leading interactions are those between
graphene and the polar modes of the SiO$_2$ substrate, and between
graphene and a possible water layer on top of the substrate. Both
effects are of similar order of magnitude in the present
approximation, where we have assumed that the water molecules are
aligned in the direction normal to the substrate.

The interactions for multi-layer graphene samples can be obtained by adding the
separate contributions from each layer.
The different dependences on distance imply that the relative strength
of the interactions  in samples with many layers can change compared
to the results of Table I. For instance, the effects of
the polar substrate $\propto z^{-2}$ and of charged impurities
$\propto z^{-1}$, which are of longer range, sum up more effectively than
the binding effect of water:
the $z^{-3}$ decay of the graphene-water interaction suggests that
only the first graphene layer
is affected by the presence of water on the substrate.
For the same reason, the presence of several  layers  of aligned water
molecules should not increase the binding,
as only the closest layer effectively contributes to the interaction
energy. On the other hand, the binding effect of water could be
enhanced if the molecules were allowed to rotate freely, therefore approaching
the high polarizability of liquid water,\cite{Setal07b}
or if they were partly ionized
by the applied field,\cite{Ketal03} leading to additional charges similar to
the Coulomb impurities present in the SiO$_2$ substrate.

It should be noted that we have considered here only long-range electrostatic
interactions, for which reliable expressions can be obtained, in
terms of well understood material parameters, like the molecular
polarizability, electric dipoles, or surface modes. Still, there is
a significant uncertainty in some parameters, like the distance of the
relevant charges to the graphene layer and the
concentration of charged impurities and water molecules.
We have not analyzed the possible
formation of chemical bonds between the carbon
atoms and the water or silanol groups at the SiO$_2$ surface.
Calculations based on the Local Density Functional
approximation\cite{Wetal07,LPP07} suggest that individual molecules
can (weakly) bind to a graphene layer with energies of $10 - 50$ meV, although
it is unclear how these estimates are changed when the molecules
interact at the same time with the graphene layer and the substrate.

\begin{table}
\begin{tabular}{||l||c|c|c||}
\hline \hline & Distance  &Dependence on &Energy \\
& (nm) &distance &  (meV \AA$^{-2}$)\\ \hline
Gate &300  &   $z^{-3} \log(z/z_s)$& $ 10^{-8}$  \\
Charged impurities &1  &$z^{-1}$ &$ 10^{-4}-10^{-2}$  \\
SiO$_2$ substrate ($z\ll l_i$) &1  & $z^{-2}$&$0.4$  \\
SiO$_2$ substrate ($z \gg l_i$) &300 nm & $z^{-3} \log(4z/l_i)$ & $10^{-8}$\\
Water molecules &0.3  &$z^{-3}$ &$1$  \\
Graphene &0.3   &$z^{-3}$ &$30$
\\ \hline \hline
\end{tabular}
\caption{Interaction energy per unit area for the mechanisms studied
in the paper. For the numerical estimates we have used typical
concentrations of $10^{10}-10^{12}$cm$^{-2}$ charged impurities and
$10^{15}$cm$^{-2}$ water molecules} \label{table}
\end{table}

\subsection{Corrugation of the graphene layer induced by the substrate.}

The attractive forces calculated in the preceding section imply that
graphene is bound to the SiO$_2$ substrate, as observed in experiments.
Our previous analysis does not
include the short range repulsive forces which determine the
equilibrium distance. We assume that the total energy near the
surface is the sum of the terms analyzed above, which have a power
law dependence on the distance, and a repulsive term, $E_{rep} ( z )
= \epsilon_{rep}  ( z_0^n / z^n )$, which also decays as a power law at long
distances, with $z_0$ an undetermined length scale. For
simplicity, we assume that the leading attractive term is due to the
presence of a water layer, which behaves as $E_{water} = - \epsilon_{w} (
z_0^3 / z^3 )$. The total energy per unit area is thus:
\begin{equation}
E ( z ) = \epsilon_{rep} \frac{z_0^n}{z^n} - \epsilon_w \frac{z_0^3}{z^3}.
\end{equation}
At the equilibrium distance, $z_{eq}$, we have:
\begin{equation}
\frac{\epsilon_{rep}}{\epsilon_w} = \frac{3}{n} \left( \frac{z_{eq}}{z_{0}}
\right)^{n-3},
\end{equation}
so that:
\begin{eqnarray}
& &E'' ( z_{eq} ) = \frac{1}{z_{eq}^2} \left[ n ( n+1 ) \epsilon_{rep} \left(
\frac{z_0}{z_{eq}} \right)^{n} - 12 \epsilon_w \left( \frac{z_0}{z_{eq}}
\right)^{3} \right] \nonumber \\
& &= 3 ( n - 3 ) \frac{\epsilon_w}{z_{eq}^2}  \left(
\frac{z_0}{z_{eq}} \right)^3 = 3 ( n - 3 ) \frac{E_{water} ( z_{eq}
)}{z_{eq}^2}.
\end{eqnarray}
Hence, the order of magnitude of the pinning potential induced by
the environment on the out of plane modes of graphene is given by $K
\propto E_{water} ( z_{eq} ) / z_{eq}^2 \sim 10^{-2}- 10^{-1}$
meV \AA$^{-4}$. Defining
the out of plane displacement as $h ( \vec{r} )$, the energy stored in a
corrugated graphene layer is:
\begin{equation}
{\cal E} \approx \frac{1}{2} \int d^2 {\vec r} \left[ \kappa   (\Delta h)^2
+ K h^2 \right],
\end{equation}
where $\kappa \approx 1$eV is the bending rigidity of
graphene.\cite{KN07,MO07} For modulations $h ( \vec{r} )$ defined by
a length scale $l$, the bending energy dominates if $l \ll l^* = (
\kappa / K )^{1/4}$, while the graphene layer can be considered
rigidly pinned to the substrate if $l \gg l^*$. Using our previous
estimates, we find $l^* \sim 10$\AA, so that the graphene layer
should follow closely the corrugations of the substrate.

The pinning by the substrate implies that the dispersion of the
flexural modes becomes:
\begin{equation}
\omega_k = \sqrt{\frac{K}{\rho} + \frac{\kappa k^4}{\rho} }
\end{equation}
where $\rho$ is the mass density of the graphene layer. At long
wavelengths, $\lim_{k \rightarrow 0} \omega_k = \omega_0 \sim
10^{-4} - 10^{-3} {\rm meV} \sim 10^{-3} - 10^{-2}$K.
\begin{figure}
\begin{center}
\includegraphics*[width=7cm]{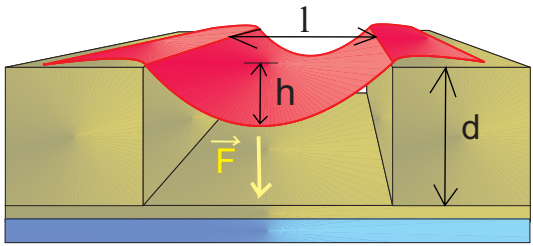}
\end{center}
\caption{Sketch of the deformation of a nanoelectromechanical device 
studied in the text.} \label{NEMs}
\end{figure}

The estimates obtained above also allow us to analyze the bending of
graphene NEMS due to the interaction with the material below, at
distance $d$.\cite{Betal07,SGN07} We assume that the lateral
dimension of the graphene cantilever is $l$, and the maximum
displacement of the graphene layer from a flat position is $h$. A
sketch of the graphene cantilever is shown in Fig.\ref{NEMs}. We
consider the force induced by charged impurities in the substrate
below the cantilever, as this is the contribution which decays more
slowly with the distance to the graphene layer (cf.  Table \ref{table}).  
If the distance of the
cantilever to the substrate is $d$, and supposing $d \gg h$,
 the gain in energy due to the
deformation of the graphene layer is 
$\Delta \mathcal{E} \sim  \epsilon_{ch} l z_0 h/ \sqrt{n_{imp}} d^2$. 
We have again
defined $z_0$ and $ \epsilon_{ch} $ by rescaling
$E_{ch}(d) = \epsilon_{ch} z_0 / d$, with
$z_0 \sim 1$ nm and
$\epsilon_{ch} \sim 10^{-4}-10^{-2}$ meV \AA$^{-2}$ depending 
on the density of impurities. The factor $l / \sqrt{n_{imp}}$ is 
included, as already mentioned  at the end of Section III.C, 
to take into account the effect of having an overall 
neutral distribution of charge in the substrate. 
This energy should compensate the elastic response
to the deformation, $\Delta \mathcal{E}_{el} \sim \kappa h^2
/ l^2$, leading to an equilibrium value: 
\begin{equation}
h \sim \frac{\epsilon_{ch} z_0}{2\kappa} \frac{l^3}{d^2 \sqrt{n_{imp}}}
\label{himp}
\end{equation}
Reminding the dependence of $\epsilon_{ch}$ on the density of impurities, 
Eq. (\ref{himp}) results in an overall behavior $h\propto \sqrt{n_{imp}}$. 
For structures such that $d\sim 300$ nm, one finds that the suspended 
graphene sheet deforms significantly (i.e. $h$ becomes comparable with $d$) 
for lenghts greater than a few $\mu$m.  

In the case of very pure substrates (i.e. neglecting the presence of 
charged impurities), the attractive force that bends the 
graphene sheet would be determined by the next corrections to the energy, 
which decay  as $z^{-3}$. 
Rewriting $E^{(2)} \sim (\epsilon_w + \epsilon_{ph} + \epsilon_{G}) (z_0/z)^3$ with $\epsilon_{ph} \simeq \epsilon_{G} \sim 0.1 meV / \AA^2$ and $\epsilon_{w} \sim 0.01 meV / \AA^2$, we see that  
those interactions are dominated by the coupling to the gate and
to the polar modes, which are of comparable magnitude.
A calculation similar to the one presented above would yield 
deformations of the order of  $h\sim  1 \AA \times (l/d)^4$, resulting even in these cases to unstable graphene sheets for  
lenghts greater than a few $\mu$m.

\subsection{Interaction with a  metallic tip in an STM
experiment}
\begin{figure}
\begin{center}
\includegraphics*[width=8cm]{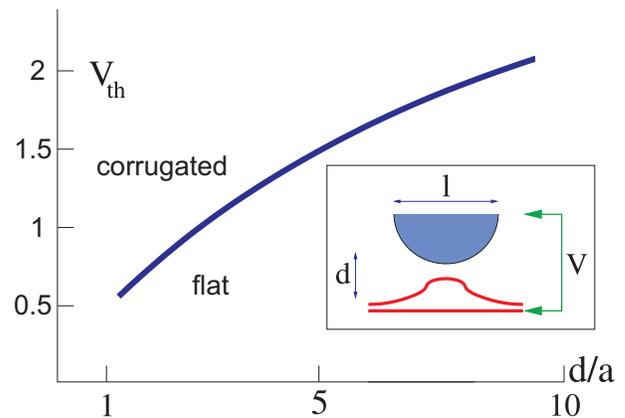}
\end{center}
\caption{Estimate of the threshold voltage as function of
graphene-tip separation needed to detach a graphene layer from the
substrate. The inset shows a sketch of the geometry considered in
the text.} \label{STM}
\end{figure}
It is known that STM tips on graphite surfaces sometimes deform the
surface graphene layer,\cite{Betal87,Setal91} and the understanding
of these deformations can be of interest for current research on
graphene.\cite{ICCFW07,Setal07,LA07}
The analysis of the electrostatic interactions between a graphene
layer and its environment allows us to estimate possible
deformations induced by an STM tip. We analyze the setup sketched in
the inset of Fig.\ref{STM}. The tip has lateral dimension $l$ and
it is located at a distance $d$ from a graphene layer. This graphene layer
interacts with an underlying substrate, and a voltage $V$ is applied
between the graphene layer and the tip. We consider three
interactions:

i) An attraction between the tip and the graphene layer, which tends
to deform the graphene, in the way shown in Fig.\ref{STM}. We
assume that this energy is purely electrostatic. A simple estimate
can be obtained by describing the setup as a capacitor where the
area of the plates is $l^2$, the distance between the plates is $d$,
and the applied voltage is $V$. The interaction energy is of the order:
\begin{equation}
\mathcal{E}_{G-tip} \approx  \frac{V^2 l^2}{8 \pi e^2 d},
\end{equation}
where we define $V$ in energy units.

 ii) The pinning of the graphene
layer to the substrate. This contribution opposes the deformation of
the layer. We write it as:
\begin{equation} \label{eq:epspin}
\mathcal{E}_{pin} \approx \epsilon_{pin} l^2,
\end{equation}
where  $\epsilon_{pin}$ is the pinning energy per
unit area. As typical values, we will use 1 meV$/$\AA$^2$ for graphene on a
water layer, and 30 meV$/$\AA$^2$ for graphene interacting with another
graphene layer, as in graphite.

iii) The rigidity of the layer against flexural deformations. This
term tends to keep the layer flat. The deformed region is likely to be (at
least) as large as the size of the STM tip. As a result, an upper bound to
the energy stored in a deformation is:
\begin{equation}
\mathcal{E}_{el} \approx  \kappa \frac{d^2}{l^2}
\end{equation}
The graphene layer will be deformed when:
\begin{equation}
\mathcal{E}_{G-tip} \gtrsim \mathcal{E}_{pin} + \mathcal{E}_{el} \label{threshold_tip}.
\end{equation}
Note that  the approximations involved in obtaining the various terms are
valid only if $d \gtrsim a$.

We consider a situation where $\epsilon_{pin} , \kappa$ and $l$
are fixed. Eq.(\ref{threshold_tip}) implies that the layer is
deformed if the voltage exceeds a threshold:
\begin{equation}
V \gtrsim V_{th} ( d ) \approx
\sqrt{8 \pi \left(\frac{\kappa e^2 d^3 }{l^4} + \epsilon_{pin}e^2 d\right)}
\end{equation}
Assuming  $l \sim 10  a$ and $d\sim a$, we see that the dominant
contribution comes from the pinning term Eq. (\ref{eq:epspin}).
Hence, in the physically relevant range $a \lesssim d \ll l$,
we can write:
\begin{equation}
V_{th} ( d ) \approx \sqrt{8 \pi \epsilon_{pin} e^2 d},
\end{equation}
independent on the tip size.
The threshold values for graphene on SiO$_2$
are of about 0.5-2 V for $d \sim 1-10$\AA,
as schematically shown in Fig.\ref{STM}.
\section{Conclusions.}

We have analyzed the electrostatic interactions between a graphene
layer and the polarizable materials which may be present in its
environment, for samples deposited on SiO$_2$. The strength of these
interactions can be obtained in terms of a few well understood
microscopic parameters, and they have a simple dependence on the
distance between the graphene layer and the system which induces the
electrostatic field. The
analysis presented here should give reliable estimates of the order
of magnitude of the different binding energies, and of their relative
strength. We have not considered the possible formation of chemical
bonds, which may alter the results when the distances between the
carbon atoms in the graphene layer and the surrounding materials is
sufficiently small.

We find that the leading effects arise from the polar modes of the
SiO$_2$ substrate, and water which may form layers on top of it.
A summary of the main results is presented in Table \ref{table}. 
The interaction
energies with systems with $N$ layers can be obtained, to a first
approximation, by adding the contributions from each layer.
The estimated magnitude of the interactions suggests that a single graphene
layer is pinned to the substrate on length scales greater than a few
lattice spacings, $\sim 10$\AA. The electrostatic binding modifies the long
wavelength, out of plane flexural modes, which acquire a finite
frequency,  $\omega_0 \sim 10^{-4} - 10^{-3}$ meV. The long range
forces considered here can also induce large deformations in
graphene NEMS. Besides, we have analyzed the possibility of deformations
of the graphene layer by an STM tip. We find that a voltage drop of
0.5 - 2 V between the tip and the sample at distances 1 - 10 \AA \,
is sufficient to deform the graphene layer.

\section{Acknowledgements.}

This work was supported by MEC (Spain) through grant
FIS2005-05478-C02-01, the Comunidad de Madrid, through the program
CITECNOMIK, CM2006-S-0505-ESP-0337, the European Union Contract
12881 (NEST) (J. S., C. S., S. F., and F. G.), MEC through grant
FIS2004-05120 and FIS2007-65723, and EU Marie Curie RTN Programme
no. MRTN-CT-2003-504574 (F.S. and J.S.). J. S. acknowledges
the I3P Program from CSIC, and C.S. the FPU Program from MEC,
for funding.  We are thankful to A. Bachtold
and A. K. Geim for many helpful insights into the relevance of water
for current experiments, to S. Vieira for useful information on
the interaction between graphene and STM tips, and to A. A. Balandin 
for a clarifying remark on our results for suspended graphene. 
\bibliography{vdW_sub}

\begin{thebibliography}{51}
\expandafter\ifx\csname natexlab\endcsname\relax\def\natexlab#1{#1}\fi
\expandafter\ifx\csname bibnamefont\endcsname\relax
  \def\bibnamefont#1{#1}\fi
\expandafter\ifx\csname bibfnamefont\endcsname\relax
  \def\bibfnamefont#1{#1}\fi
\expandafter\ifx\csname citenamefont\endcsname\relax
  \def\citenamefont#1{#1}\fi
\expandafter\ifx\csname url\endcsname\relax
  \def\url#1{\texttt{#1}}\fi
\expandafter\ifx\csname urlprefix\endcsname\relax\def\urlprefix{URL }\fi
\providecommand{\bibinfo}[2]{#2}
\providecommand{\eprint}[2][]{\url{#2}}

\bibitem[{\citenamefont{Novoselov et~al.}(2005)\citenamefont{Novoselov, Jiang,
  Schedin, Booth, Khotkevich, Morozov, and Geim}}]{Netal05b}
\bibinfo{author}{\bibfnamefont{K.~S.} \bibnamefont{Novoselov}},
  \bibinfo{author}{\bibfnamefont{D.}~\bibnamefont{Jiang}},
  \bibinfo{author}{\bibfnamefont{F.}~\bibnamefont{Schedin}},
  \bibinfo{author}{\bibfnamefont{T.~J.} \bibnamefont{Booth}},
  \bibinfo{author}{\bibfnamefont{V.~V.} \bibnamefont{Khotkevich}},
  \bibinfo{author}{\bibfnamefont{S.~V.} \bibnamefont{Morozov}},
  \bibnamefont{and} \bibinfo{author}{\bibfnamefont{A.~K.} \bibnamefont{Geim}},
  \bibinfo{journal}{Proc. Nat. Acad. Sc.} \textbf{\bibinfo{volume}{102}},
  \bibinfo{pages}{10451} (\bibinfo{year}{2005}).

\bibitem[{\citenamefont{Geim and Novoselov}(2007)}]{GN07}
\bibinfo{author}{\bibfnamefont{A.~K.} \bibnamefont{Geim}} \bibnamefont{and}
  \bibinfo{author}{\bibfnamefont{K.~S.} \bibnamefont{Novoselov}},
  \bibinfo{journal}{Nature Materials} \textbf{\bibinfo{volume}{6}},
  \bibinfo{pages}{183} (\bibinfo{year}{2007}).

\bibitem[{\citenamefont{{Castro Neto} et~al.}(2007)\citenamefont{{Castro Neto},
  Guinea, Peres, Novoselov, and Geim}}]{NGPNG07}
\bibinfo{author}{\bibfnamefont{A.~H.} \bibnamefont{{Castro Neto}}},
  \bibinfo{author}{\bibfnamefont{F.}~\bibnamefont{Guinea}},
  \bibinfo{author}{\bibfnamefont{N.~M.~R.} \bibnamefont{Peres}},
  \bibinfo{author}{\bibfnamefont{K.~S.} \bibnamefont{Novoselov}},
  \bibnamefont{and} \bibinfo{author}{\bibfnamefont{A.~K.} \bibnamefont{Geim}}
  (\bibinfo{year}{2007}), \eprint{arXiv:0709.1163}.

\bibitem[{\citenamefont{Nelson et~al.}(2004)\citenamefont{Nelson, Piran, and
  Weinberg}}]{Nelson}
\bibinfo{author}{\bibfnamefont{D.}~\bibnamefont{Nelson}},
  \bibinfo{author}{\bibfnamefont{D.}~\bibnamefont{Piran}}, \bibnamefont{and}
  \bibinfo{author}{\bibfnamefont{S.}~\bibnamefont{Weinberg}},
  \emph{\bibinfo{title}{Statistical Mechanics of Membranes and Surfaces}}
  (\bibinfo{publisher}{World Scientific (Singapore)}, \bibinfo{year}{2004}).

\bibitem[{\citenamefont{Berger et~al.}(2004)\citenamefont{Berger, Song, Li, Li,
  Ogbazghi, Feng, Dai, Marchenkov, Conrad, First et~al.}}]{Betal04}
\bibinfo{author}{\bibfnamefont{C.}~\bibnamefont{Berger}},
  \bibinfo{author}{\bibfnamefont{Z.~M.} \bibnamefont{Song}},
  \bibinfo{author}{\bibfnamefont{T.~B.} \bibnamefont{Li}},
  \bibinfo{author}{\bibfnamefont{X.~B.} \bibnamefont{Li}},
  \bibinfo{author}{\bibfnamefont{A.~Y.} \bibnamefont{Ogbazghi}},
  \bibinfo{author}{\bibfnamefont{R.}~\bibnamefont{Feng}},
  \bibinfo{author}{\bibfnamefont{Z.~T.} \bibnamefont{Dai}},
  \bibinfo{author}{\bibfnamefont{A.~N.} \bibnamefont{Marchenkov}},
  \bibinfo{author}{\bibfnamefont{E.~H.} \bibnamefont{Conrad}},
  \bibinfo{author}{\bibfnamefont{P.~N.} \bibnamefont{First}},
  \bibnamefont{et~al.}, \bibinfo{journal}{J. Phys. Chem. B}
  \textbf{\bibinfo{volume}{108}}, \bibinfo{pages}{19912}
  (\bibinfo{year}{2004}).

\bibitem[{\citenamefont{Nomura and MacDonald}(2007)}]{NM07}
\bibinfo{author}{\bibfnamefont{K.}~\bibnamefont{Nomura}} \bibnamefont{and}
  \bibinfo{author}{\bibfnamefont{A.~H.} \bibnamefont{MacDonald}},
  \bibinfo{journal}{Phys. Rev. Lett.} \textbf{\bibinfo{volume}{98}},
  \bibinfo{pages}{076602} (\bibinfo{year}{2007}).

\bibitem[{\citenamefont{Adam et~al.}(2007)\citenamefont{Adam, Hwang, Galitski,
  and {Das Sarma}}}]{AHGS07}
\bibinfo{author}{\bibfnamefont{S.}~\bibnamefont{Adam}},
  \bibinfo{author}{\bibfnamefont{E.~H.} \bibnamefont{Hwang}},
  \bibinfo{author}{\bibfnamefont{V.~M.} \bibnamefont{Galitski}},
  \bibnamefont{and} \bibinfo{author}{\bibfnamefont{S.}~\bibnamefont{{Das
  Sarma}}}, \bibinfo{journal}{Proc. Natl. Acad. Sci. USA}
  \textbf{\bibinfo{volume}{104}}, \bibinfo{pages}{18392}
  (\bibinfo{year}{2007}).

\bibitem[{\citenamefont{Schedin et~al.}(2007)\citenamefont{Schedin, Geim,
  Morozov, Jiang, Hill, Blake, and Novoselov}}]{Setal07b}
\bibinfo{author}{\bibfnamefont{F.}~\bibnamefont{Schedin}},
  \bibinfo{author}{\bibfnamefont{A.~K.} \bibnamefont{Geim}},
  \bibinfo{author}{\bibfnamefont{S.~V.} \bibnamefont{Morozov}},
  \bibinfo{author}{\bibfnamefont{D.}~\bibnamefont{Jiang}},
  \bibinfo{author}{\bibfnamefont{E.~H.} \bibnamefont{Hill}},
  \bibinfo{author}{\bibfnamefont{P.}~\bibnamefont{Blake}}, \bibnamefont{and}
  \bibinfo{author}{\bibfnamefont{K.~S.} \bibnamefont{Novoselov}},
  \bibinfo{journal}{Nature Mat.} \textbf{\bibinfo{volume}{6}},
  \bibinfo{pages}{652} (\bibinfo{year}{2007}).

\bibitem[{\citenamefont{Petrov and Rotkin}(2006)}]{PR06}
\bibinfo{author}{\bibfnamefont{A.~G.} \bibnamefont{Petrov}} \bibnamefont{and}
  \bibinfo{author}{\bibfnamefont{S.~V.} \bibnamefont{Rotkin}},
  \bibinfo{journal}{JETP Letters} \textbf{\bibinfo{volume}{84}},
  \bibinfo{pages}{156} (\bibinfo{year}{2006}).

\bibitem[{\citenamefont{Fratini and Guinea}(2007)}]{FG07}
\bibinfo{author}{\bibfnamefont{S.}~\bibnamefont{Fratini}} \bibnamefont{and}
  \bibinfo{author}{\bibfnamefont{F.}~\bibnamefont{Guinea}}
  (\bibinfo{year}{2007}), \eprint{arXiv:0711.1303}.

\bibitem[{\citenamefont{Chen et~al.}(2007)\citenamefont{Chen, Jang, Xiao,
  Ishigami, and Fuhrer}}]{CJXIF07}
\bibinfo{author}{\bibfnamefont{J.~H.} \bibnamefont{Chen}},
  \bibinfo{author}{\bibfnamefont{C.}~\bibnamefont{Jang}},
  \bibinfo{author}{\bibfnamefont{S.}~\bibnamefont{Xiao}},
  \bibinfo{author}{\bibfnamefont{M.}~\bibnamefont{Ishigami}}, \bibnamefont{and}
  \bibinfo{author}{\bibfnamefont{M.~S.} \bibnamefont{Fuhrer}}
  (\bibinfo{year}{2007}), \eprint{arXiv:0711.3646}.

\bibitem[{\citenamefont{Bolotin et~al.}(2008)\citenamefont{Bolotin, Sikes,
  Jiang, Fudenberg, Hone, Kim, and Stormer}}]{Betal08}
\bibinfo{author}{\bibfnamefont{K.~I.} \bibnamefont{Bolotin}},
  \bibinfo{author}{\bibfnamefont{K.~J.} \bibnamefont{Sikes}},
  \bibinfo{author}{\bibfnamefont{Z.}~\bibnamefont{Jiang}},
  \bibinfo{author}{\bibfnamefont{G.}~\bibnamefont{Fudenberg}},
  \bibinfo{author}{\bibfnamefont{J.}~\bibnamefont{Hone}},
  \bibinfo{author}{\bibfnamefont{P.}~\bibnamefont{Kim}}, \bibnamefont{and}
  \bibinfo{author}{\bibfnamefont{H.~L.} \bibnamefont{Stormer}}
  (\bibinfo{year}{2008}), \eprint{arXiv:0802.2389}.

\bibitem[{\citenamefont{Du et~al.}(2008)\citenamefont{Du, Skachko, Barker, and
  Andrei}}]{XuDu08}
\bibinfo{author}{\bibfnamefont{X.}~\bibnamefont{Du}},
  \bibinfo{author}{\bibfnamefont{I.}~\bibnamefont{Skachko}},
  \bibinfo{author}{\bibfnamefont{A.}~\bibnamefont{Barker}}, \bibnamefont{and}
  \bibinfo{author}{\bibfnamefont{E.~Y.} \bibnamefont{Andrei}}
  (\bibinfo{year}{2008}), \eprint{arXiv:0802.2933}.

\bibitem[{\citenamefont{Ishigami et~al.}(2007)\citenamefont{Ishigami, Chen,
  Cullen, Fuhrer, and Williams}}]{ICCFW07}
\bibinfo{author}{\bibfnamefont{M.}~\bibnamefont{Ishigami}},
  \bibinfo{author}{\bibfnamefont{J.}~\bibnamefont{Chen}},
  \bibinfo{author}{\bibfnamefont{W.}~\bibnamefont{Cullen}},
  \bibinfo{author}{\bibfnamefont{M.}~\bibnamefont{Fuhrer}}, \bibnamefont{and}
  \bibinfo{author}{\bibfnamefont{E.}~\bibnamefont{Williams}},
  \bibinfo{journal}{Nano Lett.} \textbf{\bibinfo{volume}{7}},
  \bibinfo{pages}{1643} (\bibinfo{year}{2007}).

\bibitem[{\citenamefont{Stolyarova et~al.}(2007)\citenamefont{Stolyarova, Rim,
  Ryu, Maultzsch, Kim, Brus, Heinz, Hybertsen, and Flynn}}]{Setal07}
\bibinfo{author}{\bibfnamefont{E.}~\bibnamefont{Stolyarova}},
  \bibinfo{author}{\bibfnamefont{K.~T.} \bibnamefont{Rim}},
  \bibinfo{author}{\bibfnamefont{S.}~\bibnamefont{Ryu}},
  \bibinfo{author}{\bibfnamefont{J.}~\bibnamefont{Maultzsch}},
  \bibinfo{author}{\bibfnamefont{P.}~\bibnamefont{Kim}},
  \bibinfo{author}{\bibfnamefont{L.~E.} \bibnamefont{Brus}},
  \bibinfo{author}{\bibfnamefont{T.~F.} \bibnamefont{Heinz}},
  \bibinfo{author}{\bibfnamefont{M.~S.} \bibnamefont{Hybertsen}},
  \bibnamefont{and} \bibinfo{author}{\bibfnamefont{G.~W.} \bibnamefont{Flynn}},
  \bibinfo{journal}{Proc. Natl. Acad. Sci. USA} \textbf{\bibinfo{volume}{104}},
  \bibinfo{pages}{9209} (\bibinfo{year}{2007}).

\bibitem[{\citenamefont{Bunch et~al.}(2007)\citenamefont{Bunch, van~der Zande,
  Verbridge, Frank, Tanenbaum, Parpia, Craighead, and McEuen}}]{Betal07}
\bibinfo{author}{\bibfnamefont{J.~S.} \bibnamefont{Bunch}},
  \bibinfo{author}{\bibfnamefont{A.~M.} \bibnamefont{van~der Zande}},
  \bibinfo{author}{\bibfnamefont{S.~S.} \bibnamefont{Verbridge}},
  \bibinfo{author}{\bibfnamefont{I.~W.} \bibnamefont{Frank}},
  \bibinfo{author}{\bibfnamefont{D.~M.} \bibnamefont{Tanenbaum}},
  \bibinfo{author}{\bibfnamefont{J.~M.} \bibnamefont{Parpia}},
  \bibinfo{author}{\bibfnamefont{H.~G.} \bibnamefont{Craighead}},
  \bibnamefont{and} \bibinfo{author}{\bibfnamefont{P.~L.}
  \bibnamefont{McEuen}}, \bibinfo{journal}{Science}
  \textbf{\bibinfo{volume}{315}}, \bibinfo{pages}{490} (\bibinfo{year}{2007}).

\bibitem[{\citenamefont{Katsnelson and Geim}(2007)}]{KG07}
\bibinfo{author}{\bibfnamefont{M.~I.} \bibnamefont{Katsnelson}}
  \bibnamefont{and} \bibinfo{author}{\bibfnamefont{A.~K.} \bibnamefont{Geim}}
  (\bibinfo{year}{2007}), \eprint{arXiv:0706.2490}.

\bibitem[{\citenamefont{Morozov et~al.}(2007)\citenamefont{Morozov, Novoselov,
  Katsnelson, Schedin, Elias, Jaszczak, and Geim}}]{MNKSEJG07}
\bibinfo{author}{\bibfnamefont{S.}~\bibnamefont{Morozov}},
  \bibinfo{author}{\bibfnamefont{K.~S.} \bibnamefont{Novoselov}},
  \bibinfo{author}{\bibfnamefont{M.~I.} \bibnamefont{Katsnelson}},
  \bibinfo{author}{\bibfnamefont{F.}~\bibnamefont{Schedin}},
  \bibinfo{author}{\bibfnamefont{D.}~\bibnamefont{Elias}},
  \bibinfo{author}{\bibfnamefont{J.~A.} \bibnamefont{Jaszczak}},
  \bibnamefont{and} \bibinfo{author}{\bibfnamefont{A.~K.} \bibnamefont{Geim}}
  (\bibinfo{year}{2007}), \eprint{arXiv:0710.5304}.

\bibitem[{\citenamefont{Leenaerts et~al.}(2007)\citenamefont{Leenaerts,
  Partoens, and Peeters}}]{LPP07}
\bibinfo{author}{\bibfnamefont{O.}~\bibnamefont{Leenaerts}},
  \bibinfo{author}{\bibfnamefont{B.}~\bibnamefont{Partoens}}, \bibnamefont{and}
  \bibinfo{author}{\bibfnamefont{F.~M.} \bibnamefont{Peeters}}
  (\bibinfo{year}{2007}), \eprint{arXiv:0710.1757}.

\bibitem[{\citenamefont{Wehling et~al.}(2007)\citenamefont{Wehling, Novoselov,
  Morozov, Vdovin, Katsnelson, Geim, and Lichtenstein}}]{Wetal07}
\bibinfo{author}{\bibfnamefont{T.~O.} \bibnamefont{Wehling}},
  \bibinfo{author}{\bibfnamefont{K.~S.} \bibnamefont{Novoselov}},
  \bibinfo{author}{\bibfnamefont{S.~V.} \bibnamefont{Morozov}},
  \bibinfo{author}{\bibfnamefont{E.~E.} \bibnamefont{Vdovin}},
  \bibinfo{author}{\bibfnamefont{M.~I.} \bibnamefont{Katsnelson}},
  \bibinfo{author}{\bibfnamefont{A.~K.} \bibnamefont{Geim}}, \bibnamefont{and}
  \bibinfo{author}{\bibfnamefont{A.~I.} \bibnamefont{Lichtenstein}}
  (\bibinfo{year}{2007}), \eprint{arXiv:cond-mat/0703390}.

\bibitem[{\citenamefont{Echtermeyer et~al.}(2007)\citenamefont{Echtermeyer,
  Lemme, Baus, Szafranek, Geim, and Kurz}}]{Eetal07}
\bibinfo{author}{\bibfnamefont{T.~J.} \bibnamefont{Echtermeyer}},
  \bibinfo{author}{\bibfnamefont{M.~C.} \bibnamefont{Lemme}},
  \bibinfo{author}{\bibfnamefont{M.}~\bibnamefont{Baus}},
  \bibinfo{author}{\bibfnamefont{B.~N.} \bibnamefont{Szafranek}},
  \bibinfo{author}{\bibfnamefont{A.~K.} \bibnamefont{Geim}}, \bibnamefont{and}
  \bibinfo{author}{\bibfnamefont{H.}~\bibnamefont{Kurz}}
  (\bibinfo{year}{2007}), \eprint{arXiv:0712.2026}.

\bibitem[{\citenamefont{Wu et~al.}(2007)\citenamefont{Wu, Sprinkle, X, Ming,
  Berger, and de~Heer}}]{Wetal07b}
\bibinfo{author}{\bibfnamefont{X.}~\bibnamefont{Wu}},
  \bibinfo{author}{\bibfnamefont{M.}~\bibnamefont{Sprinkle}},
  \bibinfo{author}{\bibfnamefont{L.}~\bibnamefont{X}},
  \bibinfo{author}{\bibfnamefont{F.}~\bibnamefont{Ming}},
  \bibinfo{author}{\bibfnamefont{C.}~\bibnamefont{Berger}}, \bibnamefont{and}
  \bibinfo{author}{\bibfnamefont{W.~A.} \bibnamefont{de~Heer}}
  (\bibinfo{year}{2007}), \eprint{arXiv:0712.0820}.

\bibitem[{\citenamefont{Tan and Anderson}(1983)}]{TA83}
\bibinfo{author}{\bibfnamefont{S.~L.} \bibnamefont{Tan}} \bibnamefont{and}
  \bibinfo{author}{\bibfnamefont{P.~W.} \bibnamefont{Anderson}},
  \bibinfo{journal}{Chem. Phys. Lett.} \textbf{\bibinfo{volume}{97}},
  \bibinfo{pages}{23} (\bibinfo{year}{1983}).

\bibitem[{\citenamefont{Dobson et~al.}(2006)\citenamefont{Dobson, White, and
  Rubio}}]{DWR06}
\bibinfo{author}{\bibfnamefont{J.~F.} \bibnamefont{Dobson}},
  \bibinfo{author}{\bibfnamefont{A.}~\bibnamefont{White}}, \bibnamefont{and}
  \bibinfo{author}{\bibfnamefont{A.}~\bibnamefont{Rubio}},
  \bibinfo{journal}{Phys. Rev. Lett.} \textbf{\bibinfo{volume}{96}},
  \bibinfo{pages}{073201} (\bibinfo{year}{2006}).

\bibitem[{ima()}]{image}
\bibinfo{note}{For a metal, this interaction can be written as the interaction
  between the charge distribution and its image.}

\bibitem[{\citenamefont{Gonz\'alez et~al.}(1994)\citenamefont{Gonz\'alez,
  Guinea, and Vozmediano}}]{GGV94}
\bibinfo{author}{\bibfnamefont{J.}~\bibnamefont{Gonz\'alez}},
  \bibinfo{author}{\bibfnamefont{F.}~\bibnamefont{Guinea}}, \bibnamefont{and}
  \bibinfo{author}{\bibfnamefont{M.~A.~H.} \bibnamefont{Vozmediano}},
  \bibinfo{journal}{Nucl. Phys. B} \textbf{\bibinfo{volume}{424}},
  \bibinfo{pages}{596} (\bibinfo{year}{1994}).

\bibitem[{pol()}]{polar}
\bibinfo{note}{{Note} that a polar dielectric introduces new length scales,
  $l_i = \vf / \Omega_i$ associated to the frequencies of its normal modes,
  $\Omega_i$, see next section.}

\bibitem[{\citenamefont{Wunsch et~al.}(2006)\citenamefont{Wunsch, Stauber,
  Sols, and Guinea}}]{WSSG06}
\bibinfo{author}{\bibfnamefont{B.}~\bibnamefont{Wunsch}},
  \bibinfo{author}{\bibfnamefont{T.}~\bibnamefont{Stauber}},
  \bibinfo{author}{\bibfnamefont{F.}~\bibnamefont{Sols}}, \bibnamefont{and}
  \bibinfo{author}{\bibfnamefont{F.}~\bibnamefont{Guinea}},
  \bibinfo{journal}{New J. Phys.} \textbf{\bibinfo{volume}{8}},
  \bibinfo{pages}{318} (\bibinfo{year}{2006}).

\bibitem[{\citenamefont{Guinea}(2007)}]{G07}
\bibinfo{author}{\bibfnamefont{F.}~\bibnamefont{Guinea}},
  \bibinfo{journal}{Phys. Rev. B} \textbf{\bibinfo{volume}{75}},
  \bibinfo{pages}{235433} (\bibinfo{year}{2007}).

\bibitem[{\citenamefont{Sze}(1981)}]{S81}
\bibinfo{author}{\bibfnamefont{S.}~\bibnamefont{Sze}},
  \emph{\bibinfo{title}{Physics of semiconductor devices}}
  (\bibinfo{publisher}{Wiley-Interscience (New York)}, \bibinfo{year}{1981}).

\bibitem[{\citenamefont{Guinea et~al.}(2004)\citenamefont{Guinea, Jalabert, and
  Sols}}]{GJS04}
\bibinfo{author}{\bibfnamefont{F.}~\bibnamefont{Guinea}},
  \bibinfo{author}{\bibfnamefont{R.~A.} \bibnamefont{Jalabert}},
  \bibnamefont{and} \bibinfo{author}{\bibfnamefont{F.}~\bibnamefont{Sols}},
  \bibinfo{journal}{Phys. Rev. B} \textbf{\bibinfo{volume}{70}},
  \bibinfo{pages}{085310} (\bibinfo{year}{2004}).

\bibitem[{\citenamefont{Sak}(1972)}]{S72}
\bibinfo{author}{\bibfnamefont{J.}~\bibnamefont{Sak}}, \bibinfo{journal}{Phys.
  Rev. B} \textbf{\bibinfo{volume}{6}}, \bibinfo{pages}{3981}
  (\bibinfo{year}{1972}).

\bibitem[{\citenamefont{Wang and Mahan}(1972)}]{WM72}
\bibinfo{author}{\bibfnamefont{S.~Q.} \bibnamefont{Wang}} \bibnamefont{and}
  \bibinfo{author}{\bibfnamefont{G.~D.} \bibnamefont{Mahan}},
  \textbf{\bibinfo{volume}{6}}, \bibinfo{pages}{4517} (\bibinfo{year}{1972}).

\bibitem[{\citenamefont{Mori and Ando}(1989)}]{MA89}
\bibinfo{author}{\bibfnamefont{N.}~\bibnamefont{Mori}} \bibnamefont{and}
  \bibinfo{author}{\bibfnamefont{T.}~\bibnamefont{Ando}},
  \bibinfo{journal}{Phys. Rev. B} \textbf{\bibinfo{volume}{40}},
  \bibinfo{pages}{6175} (\bibinfo{year}{1989}).

\bibitem[{\citenamefont{Hulea et~al.}(2006)\citenamefont{Hulea, Fratini, Xie,
  Mulder, Iossad, Rastelli, Ciuchi, and Morpurgo}}]{Hetal06}
\bibinfo{author}{\bibfnamefont{I.~N.} \bibnamefont{Hulea}},
  \bibinfo{author}{\bibfnamefont{S.}~\bibnamefont{Fratini}},
  \bibinfo{author}{\bibfnamefont{H.}~\bibnamefont{Xie}},
  \bibinfo{author}{\bibfnamefont{C.~L.} \bibnamefont{Mulder}},
  \bibinfo{author}{\bibfnamefont{N.~N.} \bibnamefont{Iossad}},
  \bibinfo{author}{\bibfnamefont{G.}~\bibnamefont{Rastelli}},
  \bibinfo{author}{\bibfnamefont{S.}~\bibnamefont{Ciuchi}}, \bibnamefont{and}
  \bibinfo{author}{\bibfnamefont{A.~F.} \bibnamefont{Morpurgo}},
  \bibinfo{journal}{Nature Materials} \textbf{\bibinfo{volume}{5}},
  \bibinfo{pages}{982} (\bibinfo{year}{2006}).

\bibitem[{\citenamefont{Morrow and McFarlan}(1990)}]{MM90}
\bibinfo{author}{\bibfnamefont{B.}~\bibnamefont{Morrow}} \bibnamefont{and}
  \bibinfo{author}{\bibfnamefont{A.}~\bibnamefont{McFarlan}},
  \bibinfo{journal}{J. Non-Crys. Sol.} \textbf{\bibinfo{volume}{120}},
  \bibinfo{pages}{61} (\bibinfo{year}{1990}).

\bibitem[{\citenamefont{Sneh and George}(1995)}]{SG95}
\bibinfo{author}{\bibfnamefont{O.}~\bibnamefont{Sneh}} \bibnamefont{and}
  \bibinfo{author}{\bibfnamefont{S.}~\bibnamefont{George}},
  \bibinfo{journal}{J. Phys. Chem.} \textbf{\bibinfo{volume}{99}},
  \bibinfo{pages}{4639} (\bibinfo{year}{1995}).

\bibitem[{\citenamefont{Dong et~al.}(1998)\citenamefont{Dong, Pappu, and
  Xu}}]{DPX98}
\bibinfo{author}{\bibfnamefont{Y.}~\bibnamefont{Dong}},
  \bibinfo{author}{\bibfnamefont{S.}~\bibnamefont{Pappu}}, \bibnamefont{and}
  \bibinfo{author}{\bibfnamefont{Z.}~\bibnamefont{Xu}}, \bibinfo{journal}{Anal.
  Chem.} \textbf{\bibinfo{volume}{70}}, \bibinfo{pages}{4730}
  (\bibinfo{year}{1998}).

\bibitem[{\citenamefont{Nawrocki}(1997)}]{N97}
\bibinfo{author}{\bibfnamefont{J.}~\bibnamefont{Nawrocki}},
  \bibinfo{journal}{J. Chromatogr.} \textbf{\bibinfo{volume}{779}},
  \bibinfo{pages}{29} (\bibinfo{year}{1997}).

\bibitem[{\citenamefont{{Botelho do Rego} and {Vieira Ferreira}}(2001)}]{BV01}
\bibinfo{author}{\bibfnamefont{A.~M.} \bibnamefont{{Botelho do Rego}}}
  \bibnamefont{and} \bibinfo{author}{\bibfnamefont{L.~F.} \bibnamefont{{Vieira
  Ferreira}}} (\bibinfo{publisher}{Academic Press, New York},
  \bibinfo{year}{2001}), vol.~\bibinfo{volume}{2} of
  \emph{\bibinfo{series}{Handbook of surfaces and interfaces of materials}}, p.
  \bibinfo{pages}{275}.

\bibitem[{\citenamefont{Kim et~al.}(2003)\citenamefont{Kim, Javey, Vermesh,
  Wang, Li, and Dai}}]{Ketal03}
\bibinfo{author}{\bibfnamefont{W.}~\bibnamefont{Kim}},
  \bibinfo{author}{\bibfnamefont{A.}~\bibnamefont{Javey}},
  \bibinfo{author}{\bibfnamefont{O.}~\bibnamefont{Vermesh}},
  \bibinfo{author}{\bibfnamefont{Q.}~\bibnamefont{Wang}},
  \bibinfo{author}{\bibfnamefont{Y.}~\bibnamefont{Li}}, \bibnamefont{and}
  \bibinfo{author}{\bibfnamefont{H.}~\bibnamefont{Dai}}, \bibinfo{journal}{Nano
  Lett.} \textbf{\bibinfo{volume}{3}}, \bibinfo{pages}{193}
  (\bibinfo{year}{2003}).

\bibitem[{\citenamefont{Antognozzi et~al.}(2001)\citenamefont{Antognozzi,
  Humphris, and Miles}}]{AHM01}
\bibinfo{author}{\bibfnamefont{M.}~\bibnamefont{Antognozzi}},
  \bibinfo{author}{\bibfnamefont{A.}~\bibnamefont{Humphris}}, \bibnamefont{and}
  \bibinfo{author}{\bibfnamefont{M.}~\bibnamefont{Miles}},
  \bibinfo{journal}{Appl. Phys. Lett.} \textbf{\bibinfo{volume}{78}},
  \bibinfo{pages}{300} (\bibinfo{year}{2001}).

\bibitem[{\citenamefont{Opitz et~al.}(2007)\citenamefont{Opitz, Scherge, Ahmed,
  and Schaefer}}]{OSAA07}
\bibinfo{author}{\bibfnamefont{A.}~\bibnamefont{Opitz}},
  \bibinfo{author}{\bibfnamefont{M.}~\bibnamefont{Scherge}},
  \bibinfo{author}{\bibfnamefont{S.-U.} \bibnamefont{Ahmed}}, \bibnamefont{and}
  \bibinfo{author}{\bibfnamefont{J.}~\bibnamefont{Schaefer}},
  \bibinfo{journal}{J. Appl. Phys.} \textbf{\bibinfo{volume}{101}},
  \bibinfo{pages}{064310} (\bibinfo{year}{2007}).

\bibitem[{\citenamefont{Benedict et~al.}(1998)\citenamefont{Benedict, Chopra,
  Cohen, Zettl, Louie, and Crespi}}]{Betal98}
\bibinfo{author}{\bibfnamefont{L.~X.} \bibnamefont{Benedict}},
  \bibinfo{author}{\bibfnamefont{N.~G.} \bibnamefont{Chopra}},
  \bibinfo{author}{\bibfnamefont{M.~L.} \bibnamefont{Cohen}},
  \bibinfo{author}{\bibfnamefont{A.}~\bibnamefont{Zettl}},
  \bibinfo{author}{\bibfnamefont{S.~G.} \bibnamefont{Louie}}, \bibnamefont{and}
  \bibinfo{author}{\bibfnamefont{V.~H.} \bibnamefont{Crespi}},
  \bibinfo{journal}{Chem. Phys. Lett.} \textbf{\bibinfo{volume}{286}},
  \bibinfo{pages}{490} (\bibinfo{year}{1998}).

\bibitem[{\citenamefont{Hasegawa et~al.}(2007)\citenamefont{Hasegawa,
  Nishidate, and Iyetomi}}]{HNI07}
\bibinfo{author}{\bibfnamefont{M.}~\bibnamefont{Hasegawa}},
  \bibinfo{author}{\bibfnamefont{K.}~\bibnamefont{Nishidate}},
  \bibnamefont{and} \bibinfo{author}{\bibfnamefont{H.}~\bibnamefont{Iyetomi}},
  \bibinfo{journal}{Phys. Rev. B} \textbf{\bibinfo{volume}{76}},
  \bibinfo{pages}{115424} (\bibinfo{year}{2007}).

\bibitem[{\citenamefont{Kim and {Castro Neto}}(2007)}]{KN07}
\bibinfo{author}{\bibfnamefont{E.-A.} \bibnamefont{Kim}} \bibnamefont{and}
  \bibinfo{author}{\bibfnamefont{A.~H.} \bibnamefont{{Castro Neto}}}
  (\bibinfo{year}{2007}), \eprint{arXiv:cond-mat/0702562}.

\bibitem[{\citenamefont{Marinari and {von Oppen}}(2007)}]{MO07}
\bibinfo{author}{\bibfnamefont{E.}~\bibnamefont{Marinari}} \bibnamefont{and}
  \bibinfo{author}{\bibfnamefont{F.}~\bibnamefont{{von Oppen}}}
  (\bibinfo{year}{2007}), \eprint{arXiv:0707.4350}.

\bibitem[{\citenamefont{Seo{\'a}nez et~al.}(2007)\citenamefont{Seo{\'a}nez,
  Guinea, and {Castro Neto}}}]{SGN07}
\bibinfo{author}{\bibfnamefont{C.}~\bibnamefont{Seo{\'a}nez}},
  \bibinfo{author}{\bibfnamefont{F.}~\bibnamefont{Guinea}}, \bibnamefont{and}
  \bibinfo{author}{\bibfnamefont{A.~H.} \bibnamefont{{Castro Neto}}},
  \bibinfo{journal}{Phys. Rev. B} \textbf{\bibinfo{volume}{76}},
  \bibinfo{pages}{125427} (\bibinfo{year}{2007}).

\bibitem[{\citenamefont{Batra et~al.}(1887)\citenamefont{Batra, Garc{\'\i}a,
  Rohrer, Salemink, Stoll, and Ciraci}}]{Betal87}
\bibinfo{author}{\bibfnamefont{J.~P.} \bibnamefont{Batra}},
  \bibinfo{author}{\bibfnamefont{N.}~\bibnamefont{Garc{\'\i}a}},
  \bibinfo{author}{\bibfnamefont{H.}~\bibnamefont{Rohrer}},
  \bibinfo{author}{\bibfnamefont{H.}~\bibnamefont{Salemink}},
  \bibinfo{author}{\bibfnamefont{E.}~\bibnamefont{Stoll}}, \bibnamefont{and}
  \bibinfo{author}{\bibfnamefont{S.}~\bibnamefont{Ciraci}},
  \bibinfo{journal}{Surf. Sci.} \textbf{\bibinfo{volume}{181}},
  \bibinfo{pages}{126} (\bibinfo{year}{1887}).

\bibitem[{\citenamefont{Salmeron et~al.}(1991)\citenamefont{Salmeron, Ogletree,
  Ocal, Wang, Neubauer, Kolbe, and Meyers}}]{Setal91}
\bibinfo{author}{\bibfnamefont{M.}~\bibnamefont{Salmeron}},
  \bibinfo{author}{\bibfnamefont{D.~F.} \bibnamefont{Ogletree}},
  \bibinfo{author}{\bibfnamefont{C.}~\bibnamefont{Ocal}},
  \bibinfo{author}{\bibfnamefont{H.~C.} \bibnamefont{Wang}},
  \bibinfo{author}{\bibfnamefont{G.}~\bibnamefont{Neubauer}},
  \bibinfo{author}{\bibfnamefont{W.}~\bibnamefont{Kolbe}}, \bibnamefont{and}
  \bibinfo{author}{\bibfnamefont{G.}~\bibnamefont{Meyers}},
  \bibinfo{journal}{Journ. Vac. Sci. and Tech.} \textbf{\bibinfo{volume}{9}},
  \bibinfo{pages}{1347} (\bibinfo{year}{1991}).

\bibitem[{\citenamefont{Li and Andrei}(2007)}]{LA07}
\bibinfo{author}{\bibfnamefont{G.}~\bibnamefont{Li}} \bibnamefont{and}
  \bibinfo{author}{\bibfnamefont{E.~Y.} \bibnamefont{Andrei}},
  \bibinfo{journal}{Nature Phys.} \textbf{\bibinfo{volume}{3}},
  \bibinfo{pages}{623} (\bibinfo{year}{2007}).

\end{thebibliography}
\end{document}